\documentclass[aps,pre,twocolumn,groupedaddress,showpacs]{revtex4-1}
\usepackage{graphicx}
\usepackage{dcolumn}
\usepackage{bm}
\usepackage{epsfig}
\input epsf
\epsfclipon

\begin{document}

\title{ Dynamical phase transition due to preferential cluster growth of collective emotions in online communities } 
\author{Anna Chmiel and Janusz A. Ho{\l}yst} \affiliation{Faculty of Physics, Center of Excellence for Complex Systems Research, Warsaw University of Technology, Koszykowa 75, PL-00-662 Warsaw, Poland\\ 
}
\date{\today}
\begin{abstract}

We consider a   preferential cluster growth in a one-dimensional stochastic  model describing the dynamics of a binary chain with  long-range memory. The model is driven by data corresponding to emotional patterns observed during online communities' discussions. The system undergoes a dynamical phase transition.  For low values of the preference exponent, both states are observed during the string evolution  in the majority of simulated discussion threads.  When the exponent crosses a critical value, in the majority of threads an ordered  phase emerges, i.e.  from a certain time moment only one state is represented.   The transition becomes discontinuous in the thermodynamical limit when the discussions are infinitely long and even an infinitely small preference exponent leads to the ordering behavior in every discussion thread.  Numerical simulations are in a good agreement with approximated analytical formula.

\end{abstract}  \maketitle

\section{Introduction}
It is well known  (see e.g. \cite{Lau})  that  a one-dimensional (1D)  system with short-range forces cannot undergo a phase transition at a nonzero  temperature. The situation changes when  the interaction range increases, e.g., the  Ising chain  displays a second order phase transition when spin interactions decay with the  distance $r$ as $r^{-(1+\sigma)}$  for $\sigma<1$   and non-standard critical exponents are observed for $0.5<\sigma<1$ \cite{T69}. Another example  is the 1D long-range  $q$-states Potts model in which, depending on the $\sigma$ exponent and $q$-parameter,  a first-order or a second-order phase transition is possible \cite{Bayong}.  

Some properties of  1D spatial systems with long-range interactions can be mapped  to  $N$ - step (long memory) Markov chains where transitional  probabilities depend on a system history. Analytical and numerical solutions for the resulting time-dependent probability distributions  were presented in \cite{Uk,Uks} for fixed values of the time horizon $N$. The formalism was extended   in \cite{Sur,RWR}  to an infinite-range memory that covers the whole  history  of a 1D random walker. In such  a case, a dynamical phase transition takes place from the normal diffusion to a super-diffusive behavior.   When the parameter describing  the memory influence is small enough, the variance $D_L$  of a walker position  scales with the walking time $L$ as $D_L \sim L$. It increases however as $D_L ~\sim L^\kappa $, $\kappa>1$ when  the memory influence parameter crosses a critical value. The results can explain the long-term behavior of coarse-grained DNA sequences, written texts and financial data \cite{Sur}.                    

 In this work, we consider a stochastic 1D model of preferential cluster growth where  a special form of  long-memory dynamics follows  from   recent  observations of emotional patterns in  online communities discussions \cite{Plos,hate,Frank,ja_PhyA,MariaGeorgeBosa,entropy}. In fact, complex phenomena taking place during  the information search and communication exchange  over the Internet have been investigated by several authors using diverse methods of statistical physics, see e.g.  \cite{BN,Bpre,pB,anka,Gon,human_act,jp}. The studies are facilitated by an easy access to massive data sources \cite{lazer,vesp1}. Information and opinion diffusion in online communities   is frequently compared to epidemiological phenomena \cite{in1,in2,in3,in4,in5,in5,in6,in7}. Both processes, however, need  separate approaches, what was  shown e.g.  in  recent  investigations \cite{y1,y2}  of  social contagion   in online social networks that  emerged  during a political protest in Spain.  

Our model is based on  a special collective phenomenon of emotional interactions reported  in \cite{Plos}. Consecutive comments posted on blogs, the BBC Forum, IRC channels  and  the Digg website when represented by binary variables corresponding to posts' emotional valencies \cite{em,em2,em3} tend to group in  clusters  of a  similar valence and the cluster growth rate  can be well described by a  sub-linear preferential rule \cite{Plos}.  It follows a negative comment is more likely posted  after a sequence of five negative  messages  than after four such posts. The persistent dynamics of this system has been confirmed by the Hurst exponent  analysis in \cite{Frank}. The aim of this paper is to study the global behavior of this system from the point of view of dynamical phase transitions. We will investigate when during the course of time the process of preferential cluster growth leads to the emergence of a  critical cluster that is followed by posts displaying always the same  valence and what  a fraction is  of such an ordered phase in all posts.

This paper is organized  as follows.   In Sec. \ref{i}  we describe observations of emotional clusters in massive data sets, in Sec. \ref{ii} we define a data-driven  model  for posts appearance  and in Sec. \ref{iii} we present  numerical simulations showing a   transition between a  mostly disordered (hetero-emotional) and a mostly  ordered (mono-emotional) phase  in a two-state case of such a model. The model extension to a three-state system  is studied in Sec. \ref{iiii}, and in Sec. \ref{v} we compare  critical model parameters  to  data from selected online communities.

\section{Preferential growth of emotional clusters}\label{i}

\begin{figure}
\centering
\includegraphics[width=0.8\columnwidth]{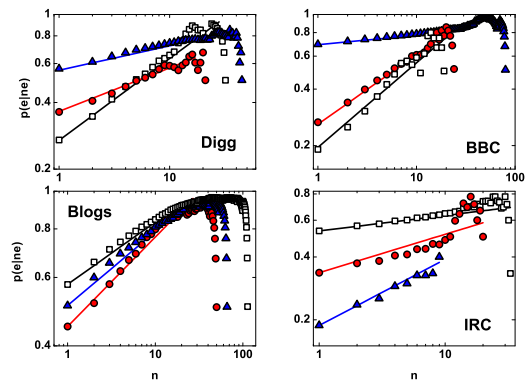}
\caption{(Color online) The conditional probability p$(e|ne)$ of the next comment occurring with the same emotion $e$ for Digg, BBC, blogs and IRC data \cite{Plos,entropy}. Symbols are data (blue triangles, red circles and white squares, for negative, positive and neutral clusters, respectively), and lines represent the fit to the preferential attraction relation $p(e|ne) = p(e|e) n^{\alpha}$}. 
\label{pn}
\end{figure}

According to the behavior found in several online communities (BBC Forum \cite{religion,world}, Digg, IRC, blog data) and presented in  \cite{Plos,entropy}, the preferential growth mechanism is the main process responsible for forming emotional clusters. It is manifested by the power-law formula for conditional probability $p(e|ne)$ that after $n$ comments with the same emotion $e$ \cite{em,em2,em3}  the next comment will express a similar sentiment. The data (see Fig. \ref{pn}) reveals the relation $p(e|ne) = p(e|e) n^{\alpha}$ where $p(e|e)$ is the conditional probability that two consecutive messages have the same emotion $e={-1,0,1}$ (negative, neutral, positive). For the description  of automatic sentiment analysis applied for the data retrieval  see \cite{Plos,mike,mike2,mike3}.  The characteristic exponent $\alpha$ represents the strength of the preferential process leading to the long-range attraction between posts of the same emotion. The probability of finding the cluster of size $n$ is proportional to the factor  $C= p(e)p(e|e)^{n-1}[(n-1)!]^{\alpha}$ responsible for appearance of the sequence of $n$ consecutive messages. It should be also taken into account that the cluster of size $n$ is defined as exactly $n$ posts with mono-emotional expressions. Thus, to get the cluster distribution function one multiplies the factor $C$  by probabilities $1-p(e)$, $1-p(e|e)n^{\alpha}$ corresponding to events  that before  and after  the cluster users write comments with emotional states  different from $e$. The analytical form of the normalization factor can be obtained only as an approximation. As a result, the  distribution of the  emotional clusters is represented  by the function: 
\begin{equation}\label{up}
P^e(n) \approx p(e|e)^{n-1}[(n-1)!]^{\alpha} [1-p(e|e)n^{\alpha}]
\end{equation}
dependent on only  two parameters $\alpha$ and $p(e|e)$.

\section{Model description}\label{ii}

Here we try to simulate the process of preferential cluster growth in an  artificial environment. To make the problem simpler, we consider a two-state system, so only positive $e=1$ or negative $e=-1$ messages can appear in this artificial discussion. Each thread has the same length $L$, not as in real data, where the thread distribution was close to a power-law relation (see Supporting Material in \cite{Plos} and  \cite{entropy}). 

The evolution rules of this  two-state system are as follows:
\begin{itemize}
\item the emotion in the  first message is randomly chosen with even probabilities $p(e=1)=p(e=-1)=1/2$  
\item the probability of emotion $e$  in the  next message  is dependent on the discussion  history. Information about this history is coded in  size $n$ of the {\bf recently} observed emotional cluster.   
The  cluster of size $n$ is defined as a sub-chain of the length $n$ of consecutive states with the same values as the valencies $e$ \cite{Plos} 
\item The process of the cluster growth is based on the behavior observed in real data. The conditional probability that the cluster containing $n$ consecutive messages with the same valency $e$ increases its length to $n+1$ is given by the equation: 
\begin{equation}\label{un}
p(e|ne)=x_e n^{\alpha_e}
\end{equation}
where $x_e$ is a constant dependent on the cluster valency $e$  (it amplifies the cluster growth, and is equivalent to $p(e|e)$) while the exponent $0 < \alpha_e < 1$ describes a strength of interactions for the emotion $e$. In the numerical simulation in each time step we randomly choose a value between  $[0;1]$. If it is  smaller than $p_e(n)$, then the cluster of the emotion $e$ is continued; otherwise, the cluster is terminated, and the opposite emotion $(-e)$ appears.
\item if $p_e(n)=1$, then the cluster reaches its {\it critical size} $n_c$, which means that starting from this moment  the discussion will be { \bf permanently ordered}  and all next messages in this thread will possess the same emotion $e$.
\end{itemize}
One can define $T_c$ as the time when the cluster of the critical size  $n_c$ appears. The  $\langle T_c \rangle$ is the average  over $R$ realizations (threads); in almost all cases  we use $R=10^{4}$.

\section{Two-state system}\label{iii}

\begin{figure}
\centering
\vskip 1.0cm
\includegraphics[width=0.8\columnwidth]{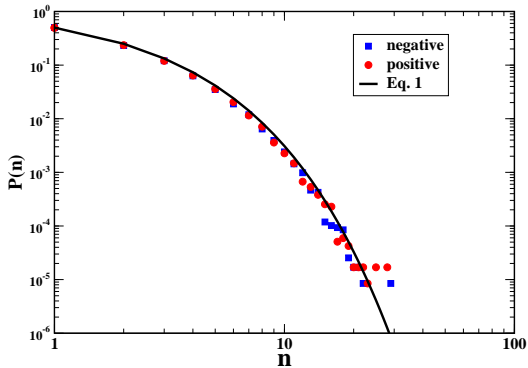}
\caption{(Color online) Numerical simulations of cluster distribution  for $x=0.5$, $\alpha=0.09$ (the same value of $\alpha$ for positive and negative emotions) $L=2 \times 10^5$.  The line corresponds to  Eq.\ref{up}.}
\label{klas}
\end{figure}

If it is not otherwise stated we shall consider  the simplest case  $x=x_1=x_{-1}=0.5$ and $\alpha_{-1}=\alpha_{1}=\alpha$. The  the probabilities of both emotions when calculated in an unordered phase (before the critical cluster occurrence)  are the same $p(-1)=p(1)=0.5$, and the  distribution of the observed cluster lengths is very similar to the one observed in the real data. In Fig. \ref{klas}, we present the cluster distribution in  artificial threads. The line comes from the theoretical prediction based on preferential cluster growth, Eq. \ref{up}.

\begin{figure}
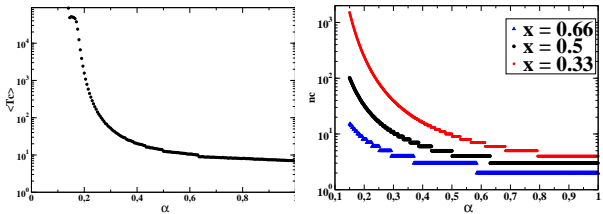

\centering
\centerline{\begin{tabular}{cc}
\includegraphics[width=0.45\columnwidth]{Tc.eps} & \includegraphics[width=0.45\columnwidth]{nc.eps}
\end{tabular}}
\caption{(Color online) Left: time $T_c$ needed for the emergence of the critical cluster for $x=0.5$, $L=10^7$. Right: size of critical cluster as a  function of $\alpha$ for $x=0.66$ (blue), $x=0.5$ (black) and $x=0.33$ (red) (from bottom to top).}
\label{1}
\end{figure}

After the transition time $T_c$, i.e, when the critical cluster appears,  the discussion  changes to the  mono-emotional thread  (MET). Starting from this moment, the   probabilities $p(-1)$ and $p(1)$ become $0$ and $1$ (or $1$ and $0$). This means that  half of the threads are nearly whole positive, and half are nearly whole negative (if the threads are long enough).
 It is obvious that the average critical time $\langle T_c \rangle$ should depend on the strength of emotional interactions, i.e, on  the exponent $\alpha$. It is also obvious that  $\langle T_c \rangle$ has to be larger or equal to the critical size of the cluster $\langle T_c \rangle \geq n_c$ (see  Fig. \ref{1}). Values of $\langle T_c \rangle$ are received from numerical simulations  and $n_c$ from Eq. \ref{un}.  

\begin{figure*}[ht]
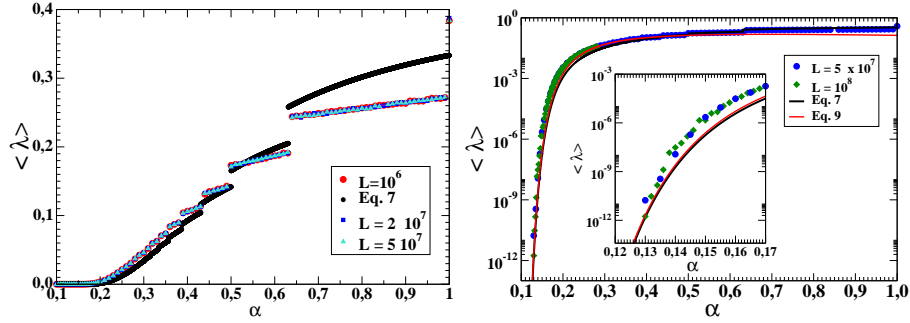


\vskip 1cm
\begin{tabular}{c c}
\epsfig{file=BBL1.eps,width=.33\textwidth} &  \epsfig{file=bbbn.eps,width=.33\textwidth} 
\end{tabular}
\caption{(Color online) Relation between the inverse of the critical time $<\lambda>$ and the exponent of affective interactions $\alpha$ for $x=0.5$ for different values of discussions lengths $L$.  Red circles: $L=10^7$, blue  squares: $L=2 \times 10^7$, sky-blue triangles: $L=5 \times 10^7$. Black circles follow   from  Eq. \ref{ddd} and are very close to the  red line from Eq.\ref{ddd2}.}
\label{2}
\end{figure*}

Since for some threads the critical cluster is not observed at all,  $\langle T_c \rangle$ is not an appropriate observable, and a more convenient variable is  a mean inverse  of the critical time  
\begin{equation}\label{un_l}
\langle \lambda \rangle=\frac{1}{\widetilde{R}}\sum_{i=1}^{i=\widetilde{R}}{\frac{1}{T_{c}^{i}}}
\end{equation}

where $\widetilde{R}$ is the number of threads that were ordered during the simulation, which means that their critical times were smaller  than the thread length. In Fig. \ref{2} we present a relation between $\langle \lambda \rangle$ and $\alpha$.  The left plot is in the linear scale and clearly displays the staircase  shape of this dependence that follows from the integer values of $T_c$ (compare Fig. \ref{1}). The right plot presents  in the   log-linear scale a rapid decrease in $\langle\lambda\rangle$ for  $\alpha \approx 0.15$. The multi-steps  shape for $\alpha > 0.3$ and a rapid decrease observed for $0.13 < \alpha < 0.2$ are only weakly  dependent on the  system size $L$. We tested this behavior for different values of $L$; for clarity, we show only representative simulations for $L=10^6$, $L=2 \times 10^7$ and $L= 5 \times 10^7$. Of course, the length of the thread $L$ influences the value $\alpha$ when the order is observed for the first time. It is  $\alpha = 0.13$ for a system of the size  $L = 5 \times 10^7$ and   $\alpha = 0.15$ when $L = 10^3$. 

Probability $P_c$ that a certain post starts a   critical cluster  can be estimated under the assumption  that in a single  discussion thread  only one critical cluster can appear
\begin{equation}\label{u8}
P_c = \langle \lambda \rangle= \frac{1}{T_c}.
\end{equation}
However, the probability of finding a cluster with the critical size can be described by a relation similar to one presented in \cite{Plos}:
\begin{equation}\label{u7}
P_c=\widetilde{P}(n_c) = A(x,\alpha)x^{n_c} [(n_c-1)!]^{\alpha},
\end{equation} 
where  $n_c=2^{\frac{1}{\alpha}}$ is the size of the critical cluster. There is a difference between Eq. \ref{u7} and an analytic calculation  presented in \cite{Plos} (see also  remarks  in Sec. \ref{i}) since here we consider  the beginning and not the end of the critical cluster.

\begin{figure}
\centering
\vskip 1cm
\includegraphics[width=0.7\columnwidth]{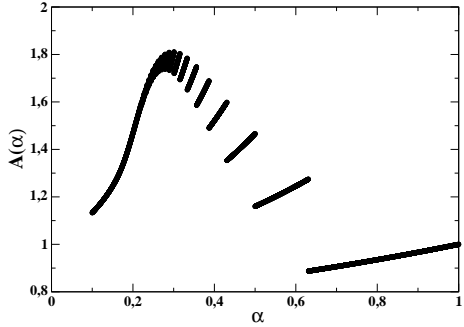}
\caption{Values of  $A(\alpha)$  estimated  for  $x=0.5$ using  {\it Mathematica}. }
\label{A}
\end{figure}

The normalization  constant in Eq. \ref{u7} 

\begin{equation}\label{sum}
A(x,\alpha)= \sum_{n=1}^{n=n_c}{ x^{n} [(n-1)!]^{\alpha}}
\end{equation}

was calculated numerically and  is  presented in Fig. \ref{A}. Since the upper limit in the above sum is $n_c$,  this normalization constant is different from that in Eq. \ref{up}.  For $\alpha \ll 1$ we get 

\begin{equation}\label{Anew}
A(x,\alpha)\approx x/(1-x).
\end{equation}

Combining Eqs. (\ref{u7})-(\ref{sum}), together we receive
\begin{equation}\label{ddd}
\langle \lambda (x,\alpha) \rangle  = A\left(x,\alpha \right) x^{x^{-1/\alpha}} \left[\left(x^{-1/\alpha} - 1\right)!\right]^{\alpha}
\end{equation}

that well fits to the behavior of $\langle \lambda(\alpha) \rangle$ received from the  numerical simulations (see the right panel in Fig. \ref{2}).  
The value of $\langle \lambda (\alpha=1) \rangle$ is not  obtained from Eq. \ref{ddd}  but  may be easily calculated from a simple branching process as:

\begin{equation}\label{un1}
\lambda(x=0.5,\alpha)=2 \sum_{n=2}^{n=n_c}{\left (\frac{1}{2}\right)^n \frac{1}{n}}=2 \ln 2-1=0.386294
\end{equation}

In the limit $\alpha \ll 1$ Eq. \ref{ddd} reduces to 

\begin{equation}\label{ddd2}
\langle \lambda (x,\alpha) \rangle \approx \frac{x^2}{1-x}\exp \left(-\alpha x^{-1/\alpha}\right)
\end{equation}
and we get $\langle \lambda (x,0) \rangle = 0$

Let us consider a discussion in thread of length $L$ with affective interactions described  by the characteristic exponent $\alpha$  and let us define a fraction of discussions that are mono-emotional  ordered (MET) from a certain moment in such a  thread as  $r(\alpha,L)=\frac{\widetilde{R}}{R}$. This value is  also a probability of the MET occurrence before time $t=L$.  It follows the value of $r$ can be written as   
\begin{equation}\label{rjh1}
r(\alpha,x,L)=1- [1-\lambda(\alpha,x)]^L
\end{equation}
 where the explicit form can be received by inserting into (\ref{rjh1})  Eqs. \ref{Anew} and \ref{ddd}. In the limit $\alpha \ll 1$ we get from (\ref{ddd2})
\begin{equation}\label{rjh2}
r(\alpha,x,L)= 1-\left [1-\frac{x^2}{1-x}\exp \left(  -\alpha x^{ (-1/\alpha)} \right) \right ]^L
\end{equation}

\begin{figure}
\centering
\vskip 1cm
\includegraphics[width=0.7\columnwidth]{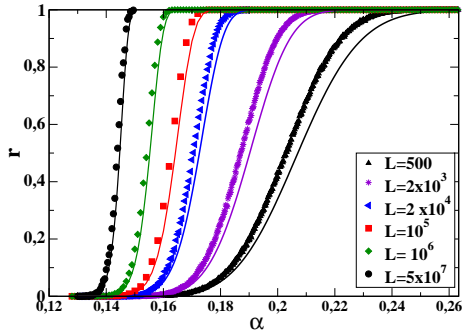}
\caption{(Color online) Fraction of  ordered threads as a function of the exponent    $\alpha$ for various thread lengths $L$. Lines correspond to  Eq. \ref{rjh2}.}
\label{h}
\end{figure}

 Results of numerical simulations  and theory from Eq. \ref{rjh2} are presented in  Fig. \ref{h}. As one could  expect a fraction $r$ of the MET phase in all threads increases with the $\alpha$ exponent and with the thread length $L$. Moreover for longer threads the agreement between Eq. \ref{rjh2} and numerical simulations is better and   the transition between the states $r\approx 0$ and $r\approx 1$   becomes steeper.  In the thermodynamical limit $L \to \infty$ this transition is discontinuous since 
\begin{equation}\label{r1}
\lim_{L \to \infty} r(\alpha=0,x,L)=0 
\end{equation}  
and  
\begin{equation}\label{r2}
\lim_{L \to \infty} r(\alpha>0,x,L)=1
\end{equation} 

\begin{figure}[ht]
\centering
\vskip 1cm
\includegraphics[width=0.7\columnwidth]{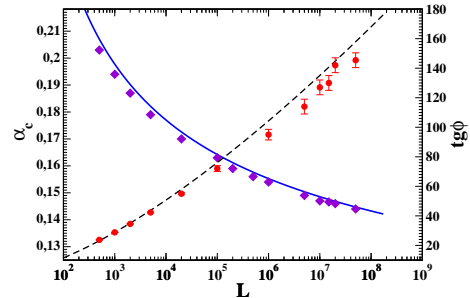}
\caption{(Color online) Dependence of  the critical value  $\alpha_c$ (violet rombs) and  the  slope of $\tan \phi $ (red circles) on the system size $L$. Solid line  corresponds to Eq. \ref{ac2} and ‪a dashed one  to Eq. \ref{tan} where the value of $\alpha_c$  was taken from the Eq. \ref{ac2}. }
\label{ac}
\end{figure}

 Let us   define the  critical value of the interaction strength  as  $\alpha_c=\alpha(r=0.5)$.
After a short algebra we get from (\ref{rjh2})  
\begin{equation} \label{rcjh1}
 1-\frac{x^2}{1-x}\exp \left[ - \alpha_c x^{(-1/\alpha_c)} \right]=2^{-\left(1/L\right )}
\end{equation}
For the  symmetrical case $x=1/2$ and  $L \gg 1$ (if it is not otherwise written we shall use these assumptions further ) we get a  simpler relation 
 \begin{equation}\label{rcjh2}
 \alpha_c 2^{(1/\alpha_c)} \approx \ln(L)-\ln [2\ln(2)] 
\end{equation} 
that can be disentangled as: 
 \begin{equation}\label{ac2}
 \alpha_c \approx - \frac{\ln (2)}{W_{-1} \left( \ln(2)/\ln(L/ \ln (4)) \right)}
\end{equation}

where  $W_{-1}(.)$ is the  lower branch of  Lambert $W$-function \cite{LamW}.   
A quantitative measure of the system behavior near the transition point $\alpha_c$ is the slope 
\begin{equation}\label{phi_def} 
\tan \phi = \left (\frac{ \partial r(\alpha,x,L)}{ \partial \alpha } \right)_{\alpha_c} 
\end{equation}

that can be expressed as :
\begin{equation}\label{tan}
 \tan \phi \approx -\frac{\ln(2)}{2} x^{-1/\alpha_c}\left[1+\frac{\ln(x)}{\alpha_c} \right].
\end{equation}
For $x=1/2$ Eq.\ref{tan} can be written as an explicit function of the length $L$ using the result (\ref{ac2}). Relations (\ref{ac2})and (\ref{tan}) are presented at  Fig.\ref{ac} where we see good fit to  corresponding numerical simulations. In the limit $L \to \infty$ the critical value $\alpha_c(L)$ tends to zero while the slope $\phi(L)$ diverges to infinity  what is  a sign of a discontinuous transition in the thermodynamical limit. It should   be stressed  that for $\alpha = 0$ the  MET phase does not exist, what is shown by Eq.\ref{r1}.

\section{Three-state system}\label{iiii}

\begin{figure}[ht]
\centering
\vskip 1cm
\includegraphics[width=0.7\columnwidth]{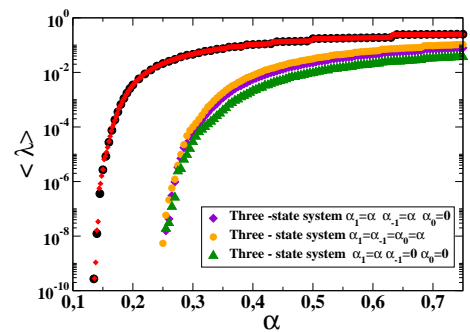}
\caption{(Color online) Relation between the observable $\langle \lambda \rangle$ and the exponent $\alpha$; black points: $x=0.5$ and $\alpha_{-1}=\alpha_{1}$ (two state-system), red diamonds: $x=0.5$ and $\alpha_{-1}=\alpha_{1}=\alpha_{0}$ (three state-system), orange, violet  and green points are for $x=0.33$ (three-state system) and different values of $\alpha$;  $L=2 \times 10^6$.}
\label{3st}
\end{figure}

A natural extension of the two-state system is to add one more state, i.e., $e\in\{-1,0,1\}$. To compare properties   of such systems  with our previous results, we considered a   symmetrical three-state model where $x_{-1}= x_0=x_1=0.5$ and $\alpha_{-1}=\alpha_{0}=\alpha_{1}$ with a  symmetrical two-state model where $x_{-1}=x_1=0.5$ and $\alpha_{-1}=\alpha_{1}$. Values of the inverse of critical time  $\langle \lambda \rangle  $ as a function of the exponent $\alpha$ are presented in  Fig. \ref{3st}. Since results  for both systems lie  on the same line, we can state that the number of possible emotional states does not influence a critical time needed for the emergence of  MET. This observation can be explained as follows. The occurrence of MET needs a growth of a critical cluster of any emotion $e$. The growth process  is dependent only on the conditional probability of cluster growth (Eq.\ref{un}) that is insensitive to the number of possible emotional states. If initial probabilities $p(e)$ of a spontaneous occurrence of every emotional state $e$ are equal and clusters of every  emotion posses the same growth parameters $\alpha_e$ and $x_e$  then an average time needed for the emergence of {\it any} critical cluster should be  independent from the number of possible emotional states.

Fig. \ref{3st}  shows the  results for an  asymmetrical three-state system where $x_{-1}= x_0=x_{1}=0.33$.  We  investigated  models   when one or two emotional states are random ($\alpha_{-1} = 0$ or/and $\alpha_0=0$) and the preferential process appears only for the remaining emotional state. We observe that for a small value of $\alpha<0.25$ all three considered curves collapsed.

\section{Real-world data}\label{v}

Let us consider the behavior of the proposed model for parameter values  corresponding to  a {\it real} exchange of messages.  For the BBC Forum, the parameters are $\alpha_{-1} = 0.051$, $\alpha_1 = 0.38$, $\alpha_0 = 0.45$ (see Ref. \cite{Plos}). In numerical simulation, the first messages were randomly chosen according to values of the emotional probabilities $p(1)=0.16$, $p(-1)=0.65$, $p(0)=0.19$ calculated for this  data set. Also the parameters $x_0=0.2$, $x_1=0.27$ and $x_{-1}=0.69$ were taken from the BBC Forum as conditional probabilities $p(e|e)$. 

It follows that the average time  corresponding to the ordering phenomenon can be estimated as   $\langle T_c^{BBC} \rangle \approx 57 000$. This value  is much larger than the average thread length observed in the BBC data. However since the BBC dataset contains in total $N_{BBC}=2,474,781$ comments \cite{Plos}, on  average there were $M_{BBC}=N_{BBC}*\lambda_{BBC}\approx 43$ cases where the MET phase could appear and  discussion participants were not able to present another emotion.   A similar situation took place for the Digg data, where $\lambda_{Digg}=9.9 \times 10^{-6}$ which corresponds to $\langle T_c^{Digg} \rangle\approx 101,000$. Since $N_{Digg}=1,646,153$ \cite{Plos}, $M_{Digg} \approx 16$. Both values $M_{BBC}$ and  $M_{Digg}$  are much lower than the total numbers of the observed threads in both communities that were correspondingly \cite{Plos} $N_{BBC}^{thread}=97,946$ and  $N_{Digg}^{thread}=129,998$. Thus although there are collective emotional interactions in above online communities, the  majority of discussions threads are not pinned to a given emotion.

\section{Conclusions}

We studied a specific long-memory stochastic process  that represents a data driven  binary  model of    emotional online discussions.  Analytical and numerical calculations show that in the course of time persistent  mono-emotional threads can emerge from the clusters of a critical size.  Such threads exist  as a majority phase above a  critical value of the emotional interactions exponent  $\alpha_c$ that value  decays to zero  when the discussion length tends to infinity. In this thermodynamical limit   there is a discontinuous transition  between a phase without mono-emotional threads and a phase when every thread is emotional ordered from a certain time moment $T_c$. The value of  $T_c$ is  independent from the system size however there are  discontinuous changes of $T_c$ for $\alpha > 0.3$. We received analytical forms for values of $T_c$, $\alpha_c$ and a fraction $r$ of the ordered threads.

 The extension of the model to a three-state dynamics does not change its main properties, e.g. the critical time $T_c$ depends in the same way on the emotional interaction exponent $\alpha$.   Applying the results of our model to the BBC and Digg data provides an evidence  that the mono-emotional  state could be present in a very small fraction  of  the observed discussion threads. 

Comparing our results to long memory Markov chains studied in \cite{Uk,Uks,Sur,RWR} we see that the preferential cluster growth process described by Eq.2 leads to a phase transition only in the thermodynamical limit $L \to \infty$. For finite systems we  observe a continuous increase  of the MET phase with the strength of interactions (see Eq. 11 and Fig 6) even for $ \alpha \to 0$. Thus our model behaves differently as compared to the $N$-step Markov model \cite{Uk,Uks,Sur,RWR} where finite size effects do not preclude a dynamical phase transition. On the other hand in the thermodynamical limit our system displays a first order phase transition that was not observed in quoted studies. 

\begin{acknowledgments}
The work was supported by EU FP7 ICT Project  {\it Collective Emotions in Cyberspace - CYBEREMOTIONS}, European COST Action MP0801 {\it Physics of Competition and Conflicts} and Polish Ministry of Science Grant 1029/7.PR UE/2009/7 and Grant  578/N-COST/2009/0. ACH wants to thank Julian Sienkiewicz for his useful comments. 
\end{acknowledgments}


\begin{thebibliography}{00}
\bibitem{em} Emotional valence is probably  the most important   of  emotion components. It stands usually at the center of emotion experience since  the relevant aspect of any 
object that elicits emotions is whether we like it or not, or whether 
it is good for us or not. For a broader description of emotional valence see \cite{em2,em3}
\end{thebibliography}
\end{document}